\documentclass{iopart}
\usepackage{graphicx}\usepackage{psfrag}
\usepackage{bm}\usepackage{epsfig}\usepackage{amssymb}
\begin{document}
\title{Vortex liquids and vortex quantum Hall states
in trapped rotating Bose gases}
\author{Uwe R. Fischer$^{1,2}$, Petr O. Fedichev$^{1,3}$, and Alessio Recati$^{1,4}$}
\address{$^{1}$Leopold-Franzens-Universit\"{a}t Innsbruck, 
Institut f\"{u}r Theoretische Physik, Technikerstrasse 25, 
A-6020 Innsbruck, Austria \\
$^2$Eberhard-Karls-Universit\"at T\"ubingen, 
Institut f\"ur Theoretische Physik, 
Auf der Morgenstelle 14, D-72076 T\"ubingen, Germany \\
$^3$Russian Research Center Kurchatov Institute, Kurchatov Square, 123182 Moscow, Russia \\
$^4$Dipartimento di Fisica, Universit\`a di Trento and INFM, I-38050 Povo, Italy}



\begin{abstract}                
We discuss 
the feasibility of quantum Hall states
of vortices in trapped low-density two-dimensional Bose gases 
with large particle interactions. For interaction strengths larger than
a critical dimensionless 2D coupling constant $g_c \approx 0.6$, 
upon increasing the rotation frequency, the system is shown to spatially 
separate into vortex lattice and melted vortex lattice (vortex liquid) phases. 
At a first critical frequency, the lattice melts completely, and
strongly correlated vortex and particle quantum Hall liquids coexist 
in inner respectively outer regions of the gas cloud. Finally, at a 
second critical frequency, the vortex liquid disappears and the strongly 
correlated particle quantum Hall state fills the whole sample. 
\end{abstract}

\maketitle


\section{Introduction}
There has been enormous interest in the last few years in 
the properties of trapped Bose gases set
under rotation, both theoretically and experimentally 
(a small selection of recent contributions is 
\cite{VortexLatticeBEC,Cornell,rosenbusch,Ho,wilkin01,sinova,erich}). 
Of particular interest is the behavior of (effectively) two-dimensional 
gases in highly correlated fractional quantum Hall states (FQHE)
\cite{laughlin2}. These states lend themselves for an exploration of 
physics beyond the mean-field picture, where the role of the strong 
magnetic field in the FQHE of ultrapure 2D electron gases is 
taken over by a fast rotation of the gas. 

The formation of FQHE states in a rapidly rotating, 
harmonically trapped dilute Bose gas, such that the
Landau level mixing due to interactions can be neglected, was studied in 
Ref.\,\cite{wilkin01}. It was found that, after the vortex lattice 
melts, at a critical filling factor of the rotational Landau levels 
the system of particles enters a highly correlated quantum ground state. 
The many-body wave function corresponding to this state was numerically 
shown to have large overlap with a bosonic version of the many-body 
quantum Hall ground state introduced by Laughlin for electrons 
\cite{laughlin2}. 

Instead of a quantum Hall fluid of {\em particles}, 
one can equally well consider the {\em vortices} in the fluid 
to be the elementary objects, forming themselves a highly 
correlated quantum fluid. Previously, a quantum Hall fluid 
of vortices has been invoked to explain certain features of 
Hall conductance and magnetization experiments in high-$T_c$ superconductors 
\cite{review,Onogi,Rozhkov,Horovitz,Sasaki}. 
In the present paper, we propose to investigate vortex liquid 
phases in a rotating, spatially inhomogeneous, 
harmonically trapped dilute Bose gas. 
It is shown, in particular, that there exists a critical
dimensionless 2D interaction strength, 
$g=g_c\approx 0.6$, above which several phases can coexist 
(we set $\hbar =  m =1$) \cite{KTBBEC}.  
In the limit of weak interaction, 
for $g<g_c$, and after the vortex lattice has melted, the particles are
confined to the lowest rotational 
Landau level and, for low enough temperatures, the 
system makes a transition into the particle Laughlin state.  
For $g> g_c$, on the other hand, 
a transition into a vortex quantum Hall state can occur. 
Upon partial lattice melting, the system separates 
into a low-density phase in which the vortex lattice has melted, and
an inner high-density region, in which a vortex lattice resides. 
At a critical rotation velocity, the lattice melts completely, and the 
inner region consists of a vortex quantum liquid, while the 
outer shell enters a highly correlated 
particle quantum Hall state.
Finally, at a second critical velocity, the vortex liquid phase disappears 
and the particle quantum Hall state fills the whole sample.  

Below, we first introduce the basic vortex Hamiltonian of superfluid hydrodynamics 
in the rotating frame, upon which our discussion of the various vortex states will be built. 
We then give in section III the criterion for the melting of the vortex lattice into a vortex 
quantum liquid for a general rotating, uncharged superfluid. In section IV, this 
criterion is specialized to harmonically trapped Bose gases, where it is demonstrated 
how the phase separation described above is obtained from the inhomogeneous density profile 
of the gas cloud. In section V, we describe possible experimental means to verify the 
existence of 
the vortex quantum liquid. Finally, the conclusions section contains a discussion of 
how the zero temperature limitation, used in the following sections, can be lifted 
at practically all  temperatures for which the vortex quantum liquid can be observed.

\section{Vortex Hamiltonian and Plasma Analogy} 
\begin{figure}[b]
\psfrag{rho}{\Large $\rho (r)$}
\psfrag{O}{\Large $\bm \Omega$}
\psfrag{Xdot}{\Large $\dot {\bm X}_i$}
\psfrag{X}{\Large ${\bm X}_i$}
\psfrag{R}{\Large $R$}
\psfrag{mv}{\Large $m_v,\kappa$}
\centerline{\epsfig{file=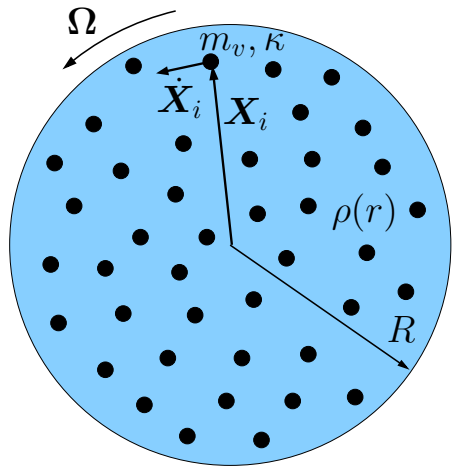,width=0.45\textwidth}}
\caption{\label{Fig1} A large collection of quantized vortices, each 
with mass $m_v$ and circulation $\kappa=2\pi$, 
in a rotating gas of radial size $R$ and 
density $\rho =\rho(r)$.
The vortices  $i$ are moving in the rotating frame at velocities 
$|\dot {\bm X}_i| \ll c_s$, at a distance $|\bm X_i|$ from the
center of rotation. This system is described by the Hamiltonian (\ref{Hv}).}
\end{figure}
Consider a two-dimensional (2D) Bose gas at temperatures well below the 
Kosterlitz-Thouless transition temperature 
$T_c^{\rm KT} = {\pi}\rho /2 $, where $\rho$ 
is the 2D particle density. 
In this superfluid state, phase fluctuations are provided by phonons and 
bound vortex-antivortex pairs.  Vorticity is a good quantum number, 
and the vortices are ``fundamental'' objects.
At sufficiently low temperatures, we can neglect the dissipation which 
arises from scattering of thermal excitations on vortices, 
so that their dynamics is conservative. The
situation considered then corresponds to the 
Magnus force dominated limit of vortex motion in very clean 
superconductors \cite{review,kolyaresonant}. 
The vortex motion is governed by a Hamiltonian corresponding to one
of point particles, with a charge equal to the quantum of 
circulation $\kappa=2\pi$, interacting with electromagnetic fields. 
For a large collection of $N_v\gg 1$ vortices (cf. Fig.\,\ref{Fig1}), 
in a frame rotating with angular velocity $\Omega$, 
the vortex Hamiltonian reads \cite{popov,annalspaper}:
\begin{eqnarray}
H_v & = & \sum_{i=1}^{N_v} \frac{\left( {\bm P}_i 
-\kappa {\bm A}_i\right)^2}{2m_v} 
-\Omega ( {\bm X}_i \times {\bm P}_i)_z \nonumber\\
& & -\frac{\rho \kappa^2}{2\pi}\sum_{i<j}^{N_v}  
\ln \left|\frac{{\bm X}_i -{\bm X}_j}{a_c}\right|. 
\label{Hv}
\end{eqnarray}
We emphasize 
that the hydrodynamic vortex Hamiltonian above does in its validity not depend on the
(microscopic) form of the many-body wave function; it describes the system
accurately if the matter we deal with is a superfluid.
The Hamiltonian is also valid if the density varies like in the (harmonically)
trapped case of currently experimentally realized Bose-Einstein condensates, 
as long as this variation takes place on scales much larger than the intervortex
distance. This is the case here, because we will consider below 
large vortex densities, with intervortex separations of order a few times 
the coherence length.

The first term in (\ref{Hv}) is the kinetic energy of vortices, 
the second stems from a $-{\bm \Omega}\cdot {\bm L}$ coupling,  
the Coriolis force on vortices, and the last
term is the repulsive logarithmic Coulomb interaction of point vortices. 
In the symmetric gauge, the vector potential $\bm A_i$, 
responsible for the vortex velocity part of the Magnus force on the vortex
(the analog of the Lorentz force), 
takes the form ${A}^a_i = \frac12\rho \epsilon^{ab} X^i_b$ ($a,b=x,y)$
\cite{annalspaper}. 
This vector potential corresponds to motion of a charged particle 
in a ``magnetic field'' perpendicular to the plane, 
of magnitude $\rho$. 
The canonical vortex momentum of the vortex $i$ is consequently given by
the sum of the kinetic and field momenta,  
\begin{equation} 
{\bm P}_i = m_v (\dot {\bm  X}_i + {\bm \Omega}\times {\bm X}_i)
+  \kappa {\bm A}_i\,.
\end{equation} 
The vortex mass is due to the density disturbance a moving vortex causes 
in the system. For a single central vortex, it is given by 
$m_v  = [\rho \kappa^2
/(4\pi c_s^{2})]  \ln [R/a_c]= [\kappa^2
/4\pi g]  \ln [R/a_c]$ \cite{duan}, where the core size 
$a_c = O(\xi_0)$ is of order the bulk coherence length $\xi_0 = 
1/\sqrt{2 g\rho}$, and $c_s=\sqrt{g\rho}$ the speed of sound 
\cite{VortexLatticewGordon}. 
The interaction strength parameter $g$ is   
related to the compressibility $\chi$ of the fluid by
$\chi = 1/g\rho^2=1/\rho c_s^2$.
We will take below the limit
of $m_v\rightarrow 0$, so that the precise value of $m_v$, in 
particular its modification in the presence of a dense vortex lattice, 
is not of importance to us, but only that $m_v$ has a finite value. 
The latter fact is important for the quantization of the vortex position and 
momentum degrees of freedom in a straightforward
canonical manner, because for $m_v=0$ 
exactly, vortex phase and configuration space coincide 
\cite{onsagerstathydro}, that is, 
vortex momenta become (gauge dependent) functions of the 
vortex coordinates alone.


The ground state of the Hamiltonian (\ref{Hv}) can be analyzed by setting all the
vortex velocities to zero (in the rotating reference frame) and finding the vortex
configuration by minimizing the energy:
\begin{equation}
H_v^{'}= \frac12 \rho\kappa\Omega \sum_{i=1}^{N_v} {\bm X}^2_i  
-\frac{\rho \kappa^2}{2\pi}\sum_{i<j}^{N_v}  
\ln \left|\frac{{\bm X}_i -{\bm X}_j}{a_c}\right|.
\end{equation}
We neglected the centrifugal potential 
$-\frac12m_v \Omega^2\sum_{i=1}^{N_v}  {\bm X}^2_i $ 
in the above expression, which is justified
by taking the small mass limit $m_v\rightarrow 0$ in (\ref{Hv}). 
In a plasma analogy, the first term in the Hamiltonian 
$H_v^{'}$ provides the neutralizing background to the potential 
created by the vortices. 
In the rotating frame the system is identical to a system of charges 
interacting by 2D Coulomb repulsion in a neutralizing background 
(the total circulation being zero as seen in the rotating frame), 
such that the ground state of the system is given by the spatially 
homogeneous vortex density $n_v = 2\Omega /\kappa =\Omega/\pi $. 

\section{Vortex Lattice Melting} 
The uniform vortex density solution is only 
a first approximation for the ground state properties of the rotating superfluid. 
Since the vortices repel each other, the energy 
of the liquid can be lowered further by forming highly correlated states.
At relatively low vortex density, the vortices form a triangular 
lattice \cite{Tkachenko}. 
The excitations above this vortex lattice ground state 
in a rotating dilute Bose-Einstein condensate have  
been the subject of recent experimental  and theoretical studies 
\cite{Coddington,GordonTkachenko}.  
They represent Tkachenko waves \cite{2ndTkachenko66}, modified 
by the interaction of the vortex lattice vibrations with sound 
\cite{GordonTkachenko}. 

According to the electrodynamical analogy,
the Abrikosov--Tkachenko vortex lattice may equivalently be understood as 
the (bosonic) Wigner crystal of vortices, analogous to the (fermionic) 
Wigner crystals formed by
electrons \cite{Wigner,Levesque,Magro}.     
On the other hand, for large vortex densities $n_v$, the lattice melts, at zero
temperature due to quantum fluctuations, and the vortices form 
a highly correlated quantum liquid. 
Indeed, long-range logarithmic interactions, experienced by our 
(bosonic) vortices, 
have already been shown by Laughlin in his seminal paper on the (fermionic)
FQHE  
to yield excellent overlap with  incompressible quantum Hall states 
(cf. \cite{laughlin2}, Table I; see Ref.\,\cite{KaneZhang} for the
present bosonic case).

In the limit of small vortex mass $m_v\rightarrow 0$, 
we have large quantum fluctuations of the vortices around their equilibrium 
lattice sites, as well as confinement of the vortices to the lowest Landau 
level of the Hamiltonian (\ref{Hv}). The dimensionless parameter 
\begin{equation}
\nu_v 
= \frac\Omega{\pi \rho} = \left(\frac{d}{\ell_v}\right)^2
,\label{nuDef}
\end{equation}
where $d\equiv \rho^{-1/2}$ 
is the interparticle distance and $\ell_v = \sqrt{\pi/\Omega}$ the mean 
distance between vortices, then controls the relative strength of quantum 
fluctuations and therefore the (zero temperature) melting 
transition into the quantum Hall state. 
The above parameter is the {\em vortex filling fraction} corresponding 
to the ratio of the number of vortices, associated with $n_v=\Omega/\pi$, and 
the number of ``flux quanta'' associated with the ``magnetic'' 
field of strength $\rho$, which is just the number of particles itself. 
The filling factor $\nu_v$ of the 
vortex rotational Landau levels is just
the inverse of the filling of particle rotational Landau levels, 
$\nu=1/\nu_v$. The Laughlin wave function for the vortices reads, in the 
usual complex notation for the vortex coordinates 
$Z_i =X_i + i Y_i$, 
\begin{equation} 
\left(\Psi_{v}\right)_{\rm Laughlin} = 
\prod_{i>j} \left( {Z_i-Z_j} \right)^{1/\nu_v} 
\exp\left[-\frac{\pi\rho}{2} \sum_i
|Z_i|^ 2\right], \label{PsiL}
\end{equation} 
where $1/\nu_v$ is even.
 
The Hamiltonian (\ref{Hv}) has been studied in the context
of vortex matter phase transitions in layered high-$T_c$ 
superconductors \cite{review,Onogi,Rozhkov,Sasaki}, where the possibility 
of a melting transition for the vortex lattice 
is due to the low Cooper-pair superfluid 
density $\rho$ in these materials \cite{review}.
In the simplest approach to melting of matter ordered in crystals, 
one assumes a Lindemann criterion \cite{Lindemann}, which
simply states that the  departure of the vortices 
from their lattice positions should approximately equal their separation 
$\ell_v$. In more refined treatments \cite{Rozhkov}, one 
takes the vortex lattice and Laughlin state wave functions for the
vortices, and evaluates their corresponding energies, equating them at
the transition point. 
In the limit of small vortex mass, the zero temperature 
transition point from the vortex
lattice to the Laughlin state is, using either method, 
Lindemann or ``exact'' evaluation of correlation energies,   
consistently within a factor of order unity found to be at 
\cite{wilkin01,sinova,Rozhkov}
\begin{equation}
\nu_v= (\nu_v)_m \approx 0.1\, \qquad (\mbox{melting filling factor}). 
\label{nuCritical}
\end{equation}
At absolute zero, 
for $\nu_v < (\nu_v)_m$, the system is in the vortex lattice phase, and
for $\nu_v > (\nu_v)_m$, it is a vortex quantum liquid, i.e., a
``melted'' vortex crystal. For small interactions $g\ll 1$, the resulting
state was shown to high accuracy to be a quantum Hall state of particles
\cite{wilkin01,sinova}. In the next section, we argue that for sufficiently
large interactions $g$, the resulting quantum Hall liquid is, conversely,
one of vortices as the elementary constituents of the liquid.

\section{Vortex Liquid States in Harmonic Traps} 
\subsection{Coexistence of Solid and Liquid Phases} 
An isotropic, harmonically trapped 2D gas possesses, in the 
Thomas-Fermi limit, 
an inhomogeneous density profile of the form 
\begin{equation}
\rho(r) =\rho_0 \left(1-\frac{r^2}{R^2}\right).
\end{equation}
Here, 
\begin{equation}
\rho_0 = \left(\frac{N}{\pi g}\right)^{1/2} \left(\omega_\perp^2 -\Omega^2 \right)^{1/2},
\qquad R^2= 2\sqrt{\frac{g N /\pi}{\omega_\perp^2 -\Omega^2}}  
\end{equation}
are the central density and 
the squared Thomas-Fermi radius of the condensate cloud, respectively. 
In contrast to an essentially
homogeneous system like a 2D superconductor, 
the melting condition (\ref{nuCritical}) then becomes
a {\em local} criterion due to its dependence on the local density. 
While the density is a locally dependent quantity, 
it is important to recognize for the ensuing analysis
that the observed vortex arrays possess a striking regularity 
\cite{VortexLatticeBEC,Cornell}. 
Even though the {\em particle} density is inhomogeneous, 
the {\em vortex} density is to a very good approximation the constant 
$n_v = \Omega/\pi$ \cite{FederClark,Muryshev,Anglin}.

To reach the vortex quantum Hall state in the 
melted phase, in preference to a particle quantum Hall state,  
the following two conditions have to be fulfilled.  
Firstly, for the lowest particle Landau level description 
to break down,  
large particle interactions have to mix the first two 
rotational particle Landau levels \cite{Ho}, 
such that $\omega_\perp +\Omega \simeq 2\Omega < g\rho $. 
Second, for melting to occur the rotation rate has to be larger than 
$\pi \rho (\nu_v)_m$. These two conditions give 
\begin{equation}
\pi \rho (\nu_v)_m < \Omega < g\rho/2. \label{Omegacond} 
\end{equation}
We conclude that, for a finite $\Omega$--region to exist  
in which the vortex quantum Hall state is preferable over other states, 
$g >g_c$  needs to be fulfilled, where 
\begin{equation}
g_c\equiv 2\pi (\nu_v)_m \approx 0.6.
\end{equation} 
In addition, our small mass assumption can only be justified 
if the Coriolis force on vortices is (much) smaller than the vortex
velocity part of the Magnus force, i.e. when $2m_v\Omega \ll \rho\kappa $.
This leads to $\Omega \ll  g\rho/\ln[R/a_c]$ which is 
for large systems more  stringent than the Landau level mixing criterion
$\Omega < g\rho/2$ by a logarithmic factor $\ln[R/a_c]/2 $. However, 
for the purpose of the present discussion, in the mesoscopic trapped gases,  
the difference between the two criteria can be neglected.  
\begin{center}
\begin{figure}[hbt]
\psfrag{OMEGA1}{$0 < \Omega < \Omega_{c1}$}
\psfrag{OMEGA2}{$\!\!\Omega_{c1} <  \Omega <  \Omega_{c2}$}
\vspace*{1em}\centerline{\epsfig{file=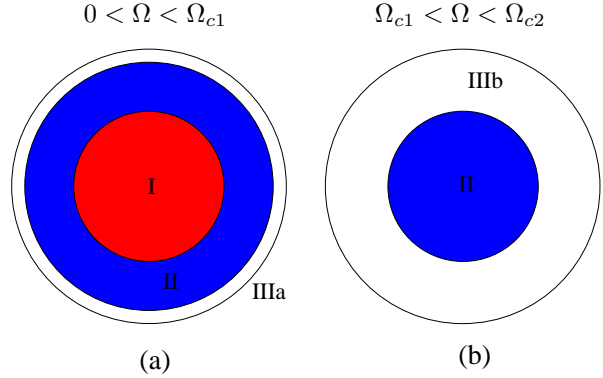,width=0.6\textwidth}}
\caption{\label{Fig2} 
Different vortex phases in a harmonically trapped gas for 
various regions of rotation velocity. (a)  
The region I signifies the vortex lattice region; 
the region II the melted lattice (vortex liquid), 
and in the very-low-density outer region IIIa the 
particles are in their lowest rotational Landau level.  
(b) For increasing  $\Omega$, the inner ``solid'' vortex lattice 
phase disappears completely at $\Omega_{c1}$ given by Eq.\,(\ref{Oc1}), 
and particle and 
vortex Laughlin state coexist in outer (IIIb) respectively inner (II) 
regions of the gas. Finally, at $\Omega> \Omega_{c2}$,
given by Eq.\,(\ref{Oc2}), the vortex quantum Hall state
disappears, and the particle quantum Hall state prevails
throughout the sample.}
\end{figure}
\vspace*{-0.35em}
\end{center}

We distinguish the following cases, cf. Fig.\,\ref{Fig2}. 
At sufficiently low angular velocity, $\Omega < \Omega_{c1}$, with  
\begin{equation}
\Delta \Omega_{c1} = \omega_\perp-\Omega_{c1} = 
\frac{g\omega_\perp}{2 \pi N [(\nu_v)_m]^2}, \label{Oc1}
\end{equation}    
the system separates into three distinct phases, cf. Fig.\,\ref{Fig2} (a)
(it is understood that all expressions for $\Delta\Omega$ 
are valid in the large $N$ limit, i.e., for $\Delta\Omega/\omega_\perp \ll 1$).
The center of the trap is occupied by the vortex lattice (region I), 
surrounded by two vortex liquid phases, in which the vortex lattice 
has melted. Phase II represents the vortex liquid, and is defined by 
Eq.\,(\ref{Omegacond}). 
The inner and outer radii of the shell are given by 
\begin{eqnarray}
\frac{x_{\rm min}^2}{R^2} & = &  1-\frac{1}{g_c\sqrt{N/g\pi}} 
\frac{2\Omega}{\sqrt{\omega_\perp^2-\Omega^2}} ,
\\
\frac{x_{\rm max}^2}{R^2} & = &  
1-\frac{1}{\sqrt{gN/\pi}} \frac{2\Omega}{\sqrt{\omega_\perp^2-\Omega^2}},
\end{eqnarray}
and the area of the shell $\Delta A$, relative to $\pi R^2$,
therefore is 
\begin{equation}
\frac{\Delta A}{\pi R^2} 
= \frac{2\Omega}{\sqrt{(gN/\pi)(\omega_\perp^2-\Omega^2)}}
\left(\frac g{g_c} -1 \right).
\end{equation} 
The number of vortices in the shell is 
\begin{eqnarray} 
(N_v)_{\rm shell} & = &  \frac{4 \Omega^2} {\omega_\perp^2 -\Omega^2} 
\left(\frac g{g_c} -1 \right).
\end{eqnarray} 
For the width of the vortex liquid shell to be a significant fraction 
of $R$ in the limit $N\rightarrow \infty$, such that 
$(N_v)_{\rm shell}\propto N$, $\Delta\Omega$ has to be a significant 
fraction of $\Delta \Omega_{c1}$. 
Furthermore, the density must be approximately homogeneous for the 
``magnetic field'' $\rho$ to be approximately constant and one well-defined 
vortex Laughlin state (\ref{PsiL}) to exist. 
The two conditions that the region II phases in Fig.\,\ref{Fig2}
be sufficiently thin for a constant density approximation to apply, 
and $(N_v)_{\rm shell}\propto  N$,   
therefore require that the coupling $g$ is sufficiently 
close to its critical value  $g_c $.  
The outermost shell IIIa shown in Fig.\,\ref{Fig2} (a) 
is filled by particles confined to their lowest
rotational Landau level.

The maximum number of vortices in the 
vortex liquid shell is reached 
at $\Omega= \Omega_{c1}$. For given $N$ and $g$, this maximum number has the value 
\begin{eqnarray} 
{\rm max}\, [(N_v)_{\rm shell}] 
= \frac{N g_c}\pi \left(1-\frac {g_c}g  \right). \label{maxNv}
\end{eqnarray}
If the rotation velocity grows and reaches $\Omega_{c1}$, the vortex lattice
completely melts, 
see Fig.\,\ref{Fig2} (b). 
The gas cloud in the region II remains superfluid. 
Further increase of $\Omega$ leads to the contraction of region II 
and eventual disappearance of the vortex liquid phase at 
$\Omega = \Omega_{c2}$, where 
\begin{equation}
\Delta\Omega_{c2} = \omega_\perp - \Omega_{c2} = 
\frac{2\pi \omega_\perp}{ g N}.  \label{Oc2}
\end{equation} 
For $\Omega_{c1}\le \Omega \le \Omega_{c2}$,  
the number of vortices in the inner melted area II is given by 
the expression
\begin{eqnarray} 
(N_v)_{\rm shell}& = &  2\Omega 
\left(\frac{gN/\pi}{\omega_\perp^2 -\Omega^2}\right)^{1/2} 
\left(1-\frac {2\sqrt\pi\Omega}{\sqrt{gN(\omega_\perp^2 -\Omega^2)}}
\right),\nonumber\\
\end{eqnarray} 
which vanishes at $\Omega =\Omega_{c2}$ and coincides with (\ref{maxNv}) 
at $\Omega =\Omega_{c1}$.

The strongly correlated fractional quantum Hall
states of particles in a harmonic trap require that the detuning 
$\Delta\Omega_{\rm particle}/\omega_\perp = 1-\Omega_{\rm particle}/\omega_\perp \approx \nu^2 g /N$ \cite{Paredes}.          
On the other hand, 
provided the inequality $g\gtrsim g_c$ is fulfilled, the vortex quantum 
Hall states occurs in an outer shell already at rotation frequencies 
$\Omega < \Omega_{c1}\sim \Omega_{\rm particle}$
smaller than the rotation frequencies 
needed for the particle quantum 
Hall states. Thus, even if only an outer shell of the sample is 
filled by the vortex quantum Hall 
state, the number of vortices can be still 
sufficiently large to observe the characteristics of the vortex
Laughlin state to be discussed in the next subsection. 

\subsection{Detection}
We now come to discuss experimental procedures to verify that the 
state of the vortex system is either crystallized or a quantum fluid. We
assume in the following that the part of the sample under consideration 
is occupied either by the vortex lattice or the vortex Laughlin phase. 

\begin{center}
\begin{figure}[t]
\vspace*{0.4em}\centerline{\epsfig{file=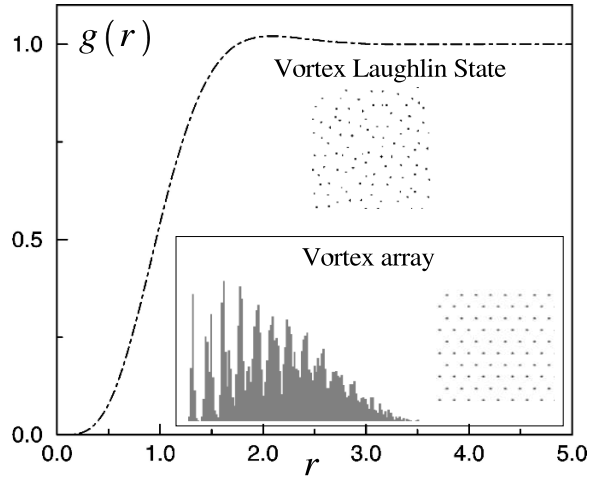,width=0.6\textwidth}}
\caption{\label{Fig3} The vortex quantum Hall state can be identified 
by measuring vortex pair-correlation 
functions in ballistic expansion pictures. The pair-correlation in the 
Laughlin state (\ref{PsiL}) is shown for $\nu_v=1/2$, from \cite{Chakraborty}. 
The two collections of points depicted represent a typical configuration in the Laughlin and 
vortex lattice states, respectively. 
The pair-correlation function of the 
lattice state shown in the inset is an 
experimentally obtained histogram, 
taken from \cite{Cornell}. It displays  
a periodic shell structure characteristic 
of crystallization in solids.}
\end{figure}
\end{center}

The melting transition can, for example, be detected by letting the gas 
expand, 
and then investigating the resulting spatial 
configuration of vortices (cf. Fig.\,\ref{Fig3}). 
The primary difference which reveals itself is 
that vortex liquids present a disordered structure compared to the
regular crystal. The information on the degree of order is encoded in the 
(assumed isotropic) pair-correlation function
$g(r)$. This information is obtained from pictures like the ones shown 
in Fig.\,\ref{Fig3}, 
by counting the number of vortices inside a shell, of radial size the intervortex spacing,  
at a distance $r$ from a given (central) vortex.  
For a crystalline ordered structure, there are oscillations 
in the pair-correlation function around its asymptotic value, 
whereas these oscillations vanish for the liquid.   
If the number of vortices is large so that the counting statistics
can be made accurate enough to extract the small-distance behavior 
$r\rightarrow 0$ of the  pair-correlation function,  
one should find $g(r) \propto r^{2/\nu_v}$ for the vortex Laughlin liquid. 
An alternative method to detect the transition to a
melted quantum Hall state was proposed in \cite{sinovaII}, by directly 
verifying a relation connecting condensate fraction and density profile.
One may also conceive of using Fourier-space techniques 
{\it in situ}, and analyze Bragg diffraction patterns by scattering light off 
a collection of vortices. 
Finally, we mention that a very recent experiment has indeed observed, 
adding a quartic trapping potential term 
to achieve $\Omega \simeq \omega_\perp$, 
a transition from a  vortex lattice to a disordered structure 
\cite{Stock}.

\bigskip
\section{Conclusions}
The preceding analysis was carried out at absolute zero.
In fact, the limitation to zero temperature 
can be relaxed. The temperature dependence of the friction force 
on the vortices can be estimated using the result of \cite{PitaFetter}; the
2D friction coefficient $\eta \sim T^4/c_s^6$. Magnus force
domination corresponds to $\rho \gg \eta$, which leads to
$T\ll \rho g^{3/4}$, provided $T\ll \rho g $. 
On the other hand, since quantum Hall states are incompressible, 
excitations of the vortex liquid 
are separated from the ground state, for finite 
$m_v$ by a gap of order 
of the Landau level spacing 
$T_{\rm Landau}= \kappa \rho / m_v \sim \rho g/\ln [R/a_c]$ 
\cite{Zhang,MikeStone}. 
In the case of large $g$, the gap is dominated by
the interaction between the vortices, and 
$T_{\rm int}\sim \rho \kappa^2 \ln [ R/a_c ]$. 
Therefore, if $T\ll T_{\rm Landau} $ respectively $T\ll T_{\rm int}$, 
excitations are frozen out and the state of the system in 
thermal equilibrium essentially coincides with the Laughlin state. 
For the interaction strengths under consideration here, 
the temperatures $T_{\rm Landau}$ and $T_{\rm int}$ 
are both comparable with the Kosterlitz-Thouless 
temperature $T_c^{\rm KT} = {\pi}\rho /2 $. 
The vortex Laughlin state thus  
forms practically for any temperature well below the 
transition temperature $T_c^{\rm KT}$ to superfluid behavior.  

To realize the vortex quantum Hall liquid, 
at the low particle densities required,  
moderately strong particle interactions are necessary, $g\gtrsim 0.6$,
both to accomodate a sufficiently large number of vortices into the rotating 
sample and to drive the system into the vortex quantum Hall state.
The ultimate limit on the maximal value of the particle coupling constant   
$g$ in a rapidly rotating, inhomogeneous 2D low-density Bose gas is not known. 
However, we may conclude from the study of Bose gases with large
scattering lengths in Ref.\,\cite{Cowell} that $g \sim O(1)$ 
should be possible before the system enters a solid, uncondensed phase.

Experimental confirmation of the vortex quantum Hall state provides 
a verification of the Hamiltonian (\ref{Hv}), which  
stems from superfluid hydrodynamics \cite{MikeStone,Giovane}, 
and the consequent quantization procedure for vortices as massive, 
``charged'' particles.
The present investigation should, then, help to shed light 
on the debate about the physical reality of vortex matter 
quantum phase transitions, whose occurence has been extensively discussed for 
the high-$T_c$ superconductors. 
It provides an analogous phenomenon in a pure, uncomplicated system, 
for which the force fields on the vortices are exactly known. 

Finally, the present investigation raises the interesting question if vortices can
also exist if only a very small number of particles per vortex, of order $5\cdots 10$, 
is present. In an abstract sense, a quantized 
vortex is a singular hole in the order parameter 
manifold with a certain residue. 
The hole effectively {\em acts} on the particles, making them spin around 
the hole in a manner prescribed by the quantization of circulation condition. 
The question which naturally poses itself, then, is if the vortex vacuum 
can also exist even in the limit of no particles being present. Formulated differently, one may ask
for the lower limit of particle density, if any, 
for which the entity quantized vortex still remains 
a well-defined physical object.

\ack
Discussions with J.\,R. Anglin, G. Blatter, J.\,I. Cirac, 
F. Dalfovo, E.\,J. Mueller, and P. Zoller are gratefully acknowledged.  
P.\,O.\,F. 
has been supported by the Austrian Science Foundation 
FWF and the 
Russian Foundation for Basic Research RFRR, 
U.\,R.\,F. by the FWF, 
and A.\,R. by the  European Union 
under Contract No. HPRN-CT-2000-00125.


\section*{References}
\begin{thebibliography}{99}
\bibitem{VortexLatticeBEC} J.\,R.  Abo-Shaeer  {\it et al.},
Science {\bf 292}, 476 (2001).
\bibitem{Cornell}
P. Engels  {\it et al.},
Phys. Rev. Lett. {\bf 89}, 100403 (2002).   
\bibitem{rosenbusch} 
P. Rosenbusch {\it et al.}, 
Phys. Rev. Lett. {\bf 88}, 250403 (2002).  
\bibitem{Ho} T.-L. Ho, Phys. Rev. Lett. {\bf 87}, 060403 (2001).  
\bibitem{wilkin01} N.\,R. Cooper, N.\,K. Wilkin, and J.\,M.\,F. Gunn, 
Phys. Rev. Lett. {\bf 87}, 120405 (2001). 
\bibitem{sinova} J. Sinova, C.\,B. Hanna, and A.\,H. MacDonald, 
Phys. Rev. Lett. {\bf 89}, 030403 (2002). 
\bibitem{erich} T.-L. Ho and E.\,J. Mueller, 
Phys. Rev. Lett. {\bf 89}, 050401 (2002).  
\bibitem{laughlin2} R.\,B. Laughlin, 
Phys. Rev. Lett. {\bf 50}, 1395 (1983).
\bibitem{review} G. Blatter, M.\,V. Feigel'man, V.\,B. Geshkenbein, 
A.\,I. Larkin, and V.\,M. Vinokur, 
Rev. Mod. Phys. {\bf 66}, 1125 (1994); contains an extensive list of 
references on vortex matter in layered high-$T_c$ materials. 
\bibitem{Onogi} T. Onogi and S. Doniach, 
Solid State Commun. {\bf 98}, 1 (1996).
\bibitem{Rozhkov} A. Rozhkov and D. Stroud, 
Phys. Rev. B {\bf 54}, R12697 (1996).
\bibitem{Horovitz} B. Horovitz, Phys. Rev. B {\bf 51}, 3989 (1995).
\bibitem{Sasaki} T. Sasaki {\it et al.}, Phys. Rev. B {\bf 57}, 10889 (1998).
\bibitem{KTBBEC}  In the limit that the 3D $s$-wave 
scattering length $a_s$ is much less than the harmonic oscillator length $l_z$ in the
perpendicular direction, the coupling constant
$g= 2\sqrt{2\pi} a_s/l_z$; 
D. S. Petrov, M. Holzmann, and G. V. Shlyapnikov, Phys. Rev. Lett.  
{\bf 84}, 2551 (2000).  
\bibitem{kolyaresonant} N.\,B. Kopnin, Phys. Rev. B {\bf 57}, 11775 (1998).
\bibitem{popov} V.\,N. Popov, Sov. Phys. JETP {\bf 37}, 341 (1973)
[Zh. \'Eksp. Th. Fiz. {\bf 64}, 672 (1973)]; 
{\em Functional Integrals and Collective Excitations} 
(Cambridge University Press, 1987).
\bibitem{annalspaper} U.\,R. Fischer, 
Ann. Phys. (N.Y.) {\bf 278}, 62 (1999).
\bibitem{duan} J.-M. Duan, Phys. Rev. B {\bf 49}, 12381 (1994).
\bibitem{VortexLatticewGordon} For simplicity, we 
do not explicitly include here the vortex core compression 
occuring for rotation frequencies $\Omega$ close to the 
perpendicular harmonic trapping $\omega_\perp$; see    
U.\,R. Fischer and G. Baym, Phys. Rev. Lett. {\bf 90}, 140402 (2003) and 
G. Baym and C.\,J. Pethick, cond-mat/0308325 
[to appear in Phys. Rev. A (2004)], 
as well as the recent experimental observation of the core compression effect 
by V. Schweikhard, I. Coddington, P. Engels, V. P. Mogendorff, and E.A. Cornell, 
Phys. Rev. Lett.  {\bf 92}, 040404 (2004).
\bibitem{onsagerstathydro} L. Onsager, 
Nuovo Cimento Suppl. {\bf 6}, 279 (1949). 
\bibitem{Coddington} I. Coddington,  P. Engels, V. Schweikhard, 
and E. A. Cornell, Phys. Rev. Lett. {\bf 91}, 100402 (2003).  
\bibitem{GordonTkachenko} G. Baym, Phys. Rev. Lett. {\bf 91}, 110402 (2003).
\bibitem{2ndTkachenko66} V.\,K. Tkachenko,  
Sov. Phys. JETP {\bf 23}, 1049 (1966)
[Zh. \'Eksp. Th. Fiz. {\bf 50}, 1573 (1966)].
\bibitem{Tkachenko} V.\,K. Tkachenko, Sov.  Phys.  JETP {\bf 22}, 1282 (1966) 
[Zh.  \'Eksp.  Teor.  Fiz.  {\bf 49}, 1875 (1965)].
\bibitem{Wigner} E. Wigner, Phys. Rev. {\bf 46}, 1002 (1934).
\bibitem{Levesque} D. Levesque, J.\,J. Weis, and A.\,H. MacDonald, 
Phys. Rev. B {\bf 30}, 1056 (1984). 
\bibitem{Magro} W.\,R. Magro and D.\,M. Ceperley, 
Phys. Rev. Lett. {\bf 73}, 826 (1994).
\bibitem{KaneZhang} C.\,L. Kane, S. Kivelson, D.\,H. Lee, and S.-C. Zhang, 
 Phys. Rev. B {\bf 43}, 3255 (1991).
\bibitem{Lindemann} F. Lindemann, Physik. Zeitschr. (Leipzig) 
{\bf 11}, 609 (1910). 
\bibitem {FederClark}  D.\,L. Feder and C.\,W. Clark, 
Phys. Rev. Lett. {\bf 87}, 190401 (2001). 
\bibitem{Muryshev} P.\,O. Fedichev and A.\,E. Muryshev, 
Phys. Rev. A {\bf 65}, 061601(R) (2002). 
\bibitem{Anglin} J.\,R. Anglin and M. Crescimanno, arXiv:cond-mat/0210063. 
\bibitem{Paredes} B. Paredes, P. Fedichev, J.\,I. Cirac, and P. Zoller,
Phys. Rev. Lett. {\bf 87}, 010402 (2001). 
\bibitem{Chakraborty} T. Chakraborty, Phys. Rev. B {\bf 57}, 8812 (1998).
\bibitem{sinovaII} J. Sinova, C.\,B. Hanna, and A.\,H. MacDonald,
Phys. Rev. Lett. {\bf 90}, 120401 (2003).   
\bibitem{Stock} V. Bretin, S. Stock, Y. Seurin, and 
J. Dalibard, 
Phys. Rev. Lett. {\bf 92}, 050403 (2004).
\bibitem{PitaFetter} L.\,P. Pitaevski\v\i, 
Sov. Phys. JETP {\bf 8}, 888 (1959); 
A.\,L. Fetter, Phys. Rev. {\bf 136}, A1488 (1964). 
\bibitem{Zhang} S.-C. Zhang, T. H. Hansson, and S. Kivelson, 
Phys. Rev. Lett. {\bf 62}, 82 (1989). 
\bibitem{MikeStone} M. Stone, Phys. Rev. B {\bf 42}, 212 (1990). 
\bibitem{Cowell} S. Cowell, H. Heiselberg, I.\,E. Mazets, J. Morales, 
V.\,R. Pandharipande, and C.\,J. Pethick,  
Phys. Rev. Lett. {\bf 88}, 210403 (2002). 
\bibitem{Giovane} S. Giovanazzi, L. Pitaevski\v\i\/, and S. Stringari, 
Phys. Rev. Lett. {\bf 72}, 3230 (1994).
\end{thebibliography}
\end{document}